\documentclass{article} 
\usepackage{iclr2026_malgai,times}

\usepackage{amsmath,amsfonts,bm}

\def\eqref#1{equation~\ref{#1}}

\def\1{\bm{1}}

\DeclareMathAlphabet{\mathsfit}{\encodingdefault}{\sfdefault}{m}{sl}
\SetMathAlphabet{\mathsfit}{bold}{\encodingdefault}{\sfdefault}{bx}{n}

\usepackage{enumitem}
\usepackage{hyperref}
\usepackage{url}
\usepackage{booktabs}
\usepackage{multirow}
\usepackage{float}
\usepackage{graphicx}
\usepackage{subcaption}
\usepackage[skip=2pt]{caption}
\usepackage{amssymb} 
\usepackage{pifont}  
\usepackage{xcolor}  
\usepackage[most]{tcolorbox}
\usepackage{pifont}
\usepackage{tcolorbox}
\usepackage{amssymb}      
\tcbuselibrary{skins, breakable}

\newcommand{\cmark}{\textcolor{green!60!black}{\checkmark}} 
\newcommand{\xmark}{\textcolor{red}{\ding{55}}}             

\title{Can Small Agents Collaborate \\ to Beat a Single Large Language Model?}

\author{%
Agata Żywot$^{1}$,
Xinyi Chen$^{1}$,
Yifei Yuan$^{2}$,
Anders Søgaard$^{3}$,
Maarten de Rijke$^{1}$ \\
$^{1}$University of Amsterdam, 
$^{2}$ETH Zürich,
$^{3}$University of Copenhagen\\
\texttt{\{a.zywot,x.chen,m.derijke\}@uva.nl}, 
\texttt{y.yuan@ethz.ch}, 
\texttt{soegaard@di.ku.dk}
}

\iclrfinalcopy 
\begin{document}

\maketitle

\begin{abstract}

Recent progress in language modeling has largely relied on scaling model size, yet larger models do not reliably improve performance on tasks requiring multi-step reasoning and tool use. Multi-agent collaboration offers a potential alternative, raising a key question: can well-organized systems built from smaller models outperform much larger language models? We address this question using a minimally designed multi-agent system with a single orchestrator and a small set of specialized sub-agents with restricted communication. On tool-intensive benchmarks spanning factual retrieval, multi-hop reasoning, scientific question answering, and mathematical problem solving, we conduct controlled comparisons between small multi-agent systems and large single-agent models. We find that small multi-agent systems can outperform substantially larger single-agent models, even when the latter have direct access to tools. Reasoning at the orchestrator yields the largest gains, while enabling reasoning in sub-agents provides limited or negative benefits. Overall system performance is driven primarily by orchestrator capacity rather than sub-agent capacity. These results suggest that improved agentic performance depends more on architectural orchestration than on raw model scaling.

\end{abstract}

\section{Introduction}
Recent progress in language modeling has been driven largely by scaling model size, yielding impressive gains across a wide range of benchmarks. However, scaling a single monolithic model comes with substantial computational cost and does not always translate into reliable performance on tasks that require multi-step reasoning, tool use, and coordination across heterogeneous information sources \citep{shojaee2025illusion, chen2024sifo, schick2023toolformer}. In such settings, even very large language models (LLMs) may struggle to decompose complex problems, select appropriate tools, or consistently enforce hard constraints~\citep{li2026single}. 

Multi-agent systems (MAS) offer a promising alternative to monolithic scaling by distributing responsibilities across multiple specialized components. Instead of relying on a single model to perform planning, reasoning, tool selection, and execution, MAS separate these roles across agents with complementary capabilities~\citep{wu2025agenticreasoningstreamlinedframework, hong2023metagpt, go2025lira}. While recent work has shown encouraging results for agentic approaches, existing systems are often complex, rely on large models throughout, or lack controlled analyses that isolate which design choices actually drive performance. As a result, it remains unclear when small-model collaboration can outperform large single-agent models, where explicit reasoning mechanisms should be applied, and which components of a multi-agent system matter most.

In this work, we investigate these questions with a simple orchestration multi-agent system setup. Our system consists of a single orchestrator responsible for planning and coordination, together with a set of specialized sub-agents equipped with distinct tools or model capabilities. By deliberately restricting communication to occur only between the orchestrator and individual sub-agents, we separate planning from execution and avoid unnecessary coordination complexity. This minimal design allows us to systematically vary model size, reasoning mechanisms, and component roles while keeping the overall architecture simple and interpretable.

\newpage
Using a diverse set of agentic and reasoning benchmarks---GAIA 
\citep{mialon2023gaiabenchmarkgeneralai}, GPQA \citep{rein2023gpqagraduatelevelgoogleproofqa}, AIME, 
MuSiQue \citep{trivedi2022musiquemultihopquestionssinglehop}, and HLE \citep{phan2025lastexam}---as 
a testbed, we study the following research questions:
\begin{itemize}[leftmargin=*]
    \setlength{\itemsep}{0pt}
    \item \textbf{RQ1:} Can a multi-agent system of small models outperform a larger single-agent model?
    \item \textbf{RQ2:} How does explicit thinking affect performance at different levels of the hierarchy?
    \item \textbf{RQ3:} Is the performance driven by the Orchestrator's capacity or the Sub-agents' size?
\end{itemize}

Our empirical results show that small multi-agent systems can outperform substantially larger single-agent models, even when the latter are equipped with direct tool use. We further find that the benefits of explicit thinking are yields the largest gains when applied at the orchestrator level, while enabling thinking in sub-agents provides limited or even negative benefits. Finally, we demonstrate that system performance is primarily driven by orchestrator capacity rather than sub-agent capacity, suggesting that multi-agent systems are planner-limited rather than executor-limited.

\textbf{Contributions.} This paper makes three contributions: (1) We introduce a multi-agent system with minimal design that enables 
controlled comparisons between small collaborative models and large 
single-agent LLMs across five benchmarks covering factual 
retrieval, multi-hop reasoning, scientific question answering, and 
mathematical reasoning. (2) We conduct a systematic analysis of explicit thinking in multi-agent systems, isolating its effects across different components. (3) We analyze the roles and capacities of the orchestrator and sub-agents, and quantify their influence on overall system performance. Together, these findings show that small-model collaboration can outperform monolithic large models when organized effectively, and that performance gains depend more on system design, particularly orchestration and reasoning placement, than on model scale alone.

\section{Related Work}
\paragraph{Tool-augmented language models.}
A growing body of work demonstrates that external tools significantly enhance the problem-solving abilities of language models. Systems integrating web search, code execution, or retrieval mechanisms enable models to overcome knowledge cutoffs and computational limitations. Frameworks such as ReAct \citep{yao2023reactsynergizingreasoningacting} and the \textit{Agentic-Reasoning} framework \citep{wu2025agenticreasoningstreamlinedframework} exemplify this approach by decomposing reasoning into planning and tool-invocation stages. While prior work shows strong gains for large models, it remains unclear to\,what extent such gains transfer to smaller backbones.

\paragraph{Multi-agent collaboration.}
Multi-agent systems distribute reasoning across specialized components or roles~\citep{wangopenhands, phan2024hyperagent, hong2023metagpt, go2025lira}. Approaches such as ToolOrchestra \citep{su2025toolorchestraelevatingintelligenceefficient} and MaAS \citep{zhang2025multiagentarchitecturesearchagentic} show that collaboration can outperform monolithic systems while reducing cost. However, most existing work focuses on architectural search, leaving open how collaboration interacts with explicit thinking and tool usage across model sizes. Our work directly addresses this gap.

\paragraph{Explicit thinking.}
Explicit reasoning mechanisms, such as Chain-of-Thought (CoT) prompting and planner-style decomposition, have been shown to improve performance on arithmetic and symbolic reasoning benchmarks by encouraging step-by-step problem decomposition \citep{guo2025deepseek, yang2025qwen3technicalreport, wei2023chainofthoughtpromptingelicitsreasoning}. However, recent work highlights important limitations of explicit thinking. CoT prompting can obscure hallucination cues and complicate model behavior, e.g., leading models to overcommit to\,intermediate steps that are incorrect \citep{cheng2025chainofthought}. Moreover, explicit reasoning does not universally improve performance and can even degrade accuracy by introducing unnecessary or noisy intermediate steps, particularly when model capacity is limited \citep{zheng2025cursecot}. As\,a\,result, when and where explicit thinking is beneficial, especially in tool-augmented agentic settings, remains an open research question.

\section{Agentic Pipeline Design}

\begin{figure}[h]
    \centering
    \includegraphics[width=\textwidth]{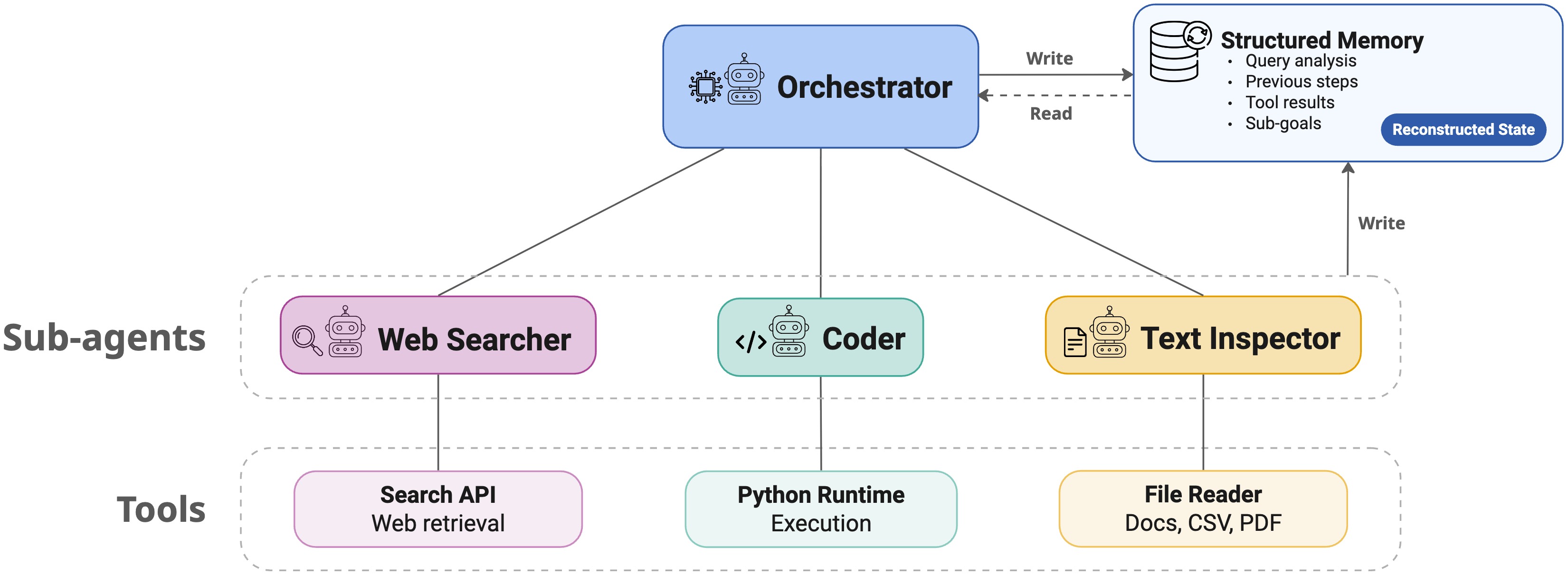}
    \caption{Overview of the Multi-Agent Orchestration Framework: The architecture decouples high-level planning (\textit{Orchestrator}) from specialized execution (\textit{Sub-agents}). Each expert sub-agent manages a specific tool domain, enabling modular task decomposition.}
    \label{fig:framework_diagram}
\end{figure}

We adopt a minimal orchestration design following \citet{wu2025agenticreasoningstreamlinedframework, li2025intheflowagenticoptimizationeffective}. As shown in Figure~\ref{fig:framework_diagram}, the system consists of three components: an orchestrator, a small set of specialized sub-agents, and a shared structured memory.

\paragraph{Orchestrator.} The orchestrator serves as the central control component. Upon receiving a user query, the orchestrator performs an initial analysis that produces a task decomposition, selects appropriate tools, and outlines an execution strategy. It then dispatches sub-tasks to the relevant sub-agents and controls the agent loop, collecting outputs, updating state, and deciding whether to issue further sub-tasks or synthesize a final response. 

\paragraph{Sub-agents.} Each sub-agent is specialized for a particular capability through access to domain-specific tools (e.g., web search, code execution). To keep the system simple and enforce a clear separation between planning and execution, communication is restricted to occur only between the orchestrator and individual sub-agents; sub-agents do not communicate with one another. This design avoids complex coordination patterns among sub-agents and places responsibility for global reasoning and decision-making entirely on the orchestrator. We implement three sub-agents covering the core capabilities required for complex multi-step reasoning.
\begin{itemize}[leftmargin=*]
    \item \textbf{Web Searcher.} Retrieves external information via a search API, providing access to up-to-date factual ontent beyond the model's parametric knowledge.
    \item \textbf{Coder.} Executes Python code in a sandboxed environment for numerical computation, algorithmic reasoning,
    and result verification.
    \item \textbf{File Inspector.} Parses external file formats, including plain text, CSV, and PDF, and provides structured access to their content for downstream reasoning.
\end{itemize}

\paragraph{Structured memory.} Rather than passing full interaction histories between components, the system maintains a shared structured memory store inspired by \citet{li2025intheflowagenticoptimizationeffective}. This store holds key intermediate results produced during task execution, organized into four categories: query analysis, previous steps, tool results, and sub-goals. Both the orchestrator and sub-agents write to this store, but only the orchestrator reads from it. At each orchestration step, the orchestrator reconstructs the current task state from memory, referred to as the \textit{Reconstructed State}, enabling coherent multi-step reasoning without requiring agents to maintain their own context across turns.

\section{Experimental Setup}

\paragraph{Models.}
We evaluate Qwen3 models \citep{yang2025qwen3technicalreport} ranging from 1.7B to 32B parameters to assess the impact of model scale on orchestration and execution. Specifically, we utilize the 8B and 32B models as Orchestrators. We also introduce smaller 1.7B models as sub-agents to test whether lightweight experts can effectively support a larger orchestrator.

\paragraph{Tool usage configurations.} To measure the effect of multi-agent coordination independently of tool access, we compare two configurations. In the \textbf{Sub-agent} configuration (our default), the orchestrator delegates tool calls to specialized sub-agents and maintains structured memory as described above. In the \textbf{Direct} configuration, the orchestrator connects to all tools itself and handles both planning and execution without sub-agents or structured memory. Comparing the two isolates the contribution of modular task delegation from the contribution of tool access alone.

\paragraph{Thinking configuration.}  Explicit reasoning can be enabled independently at different levels of the hierarchy, raising two questions: whether thinking improves performance, and where in the hierarchy it matters most. We evaluate four configurations: \textbf{None}, where no component produces explicit thinking traces; \textbf{Sub-agent only}, where thinking is enabled only for the sub-agents; \textbf{Orchestrator only}, where thinking is enabled only for the orchestrator; and \textbf{All}, where thinking is enabled for both. 

\paragraph{Datasets.} We evaluate on five benchmarks spanning the core demands of agentic and reasoning systems: GAIA \citep{mialon2023gaiabenchmarkgeneralai} for end-to-end tool use, GPQA \citep{rein2023gpqagraduatelevelgoogleproofqa} and HLE \citep{phan2025lastexam} for expert-level scientific reasoning, AIME \citep{aime} for mathematical problem-solving, and MuSiQue \citep{trivedi2022musiquemultihopquestionssinglehop} for multi-hop retrieval. We provide detailed descriptions of each benchmark in Appendix~\ref{app:benchmarks}.

\paragraph{Evaluation metrics.} 

We report accuracy across all benchmarks as the primary evaluation metric,
directly reflecting whether an agentic system successfully completes a task
end-to-end. For free-form numerical and mathematical answers, we use a
mathematical verifier to assess correctness. This consistent focus on
end-to-end accuracy aligns with our research questions on practical agentic
capability rather than intermediate reasoning quality.

\section{Results}
\subsection{Small Multi-Agent Systems vs. Large Single-Agent Models}
\label{sec:main_results}

\begin{table}[h]
    \centering
    \caption{Comparison of base models, single-agent direct tool use, and our multi-agent 
framework across reasoning benchmarks. Values in parentheses denote the accuracy change 
relative to the Qwen3-8B baseline without tools or explicit thinking. With orchestrator 
reasoning, the 8B multi-agent system becomes competitive with a 32B single-agent model 
with direct tool access, highlighting the importance of orchestration beyond model scale.}
    \includegraphics[width=0.95\textwidth]{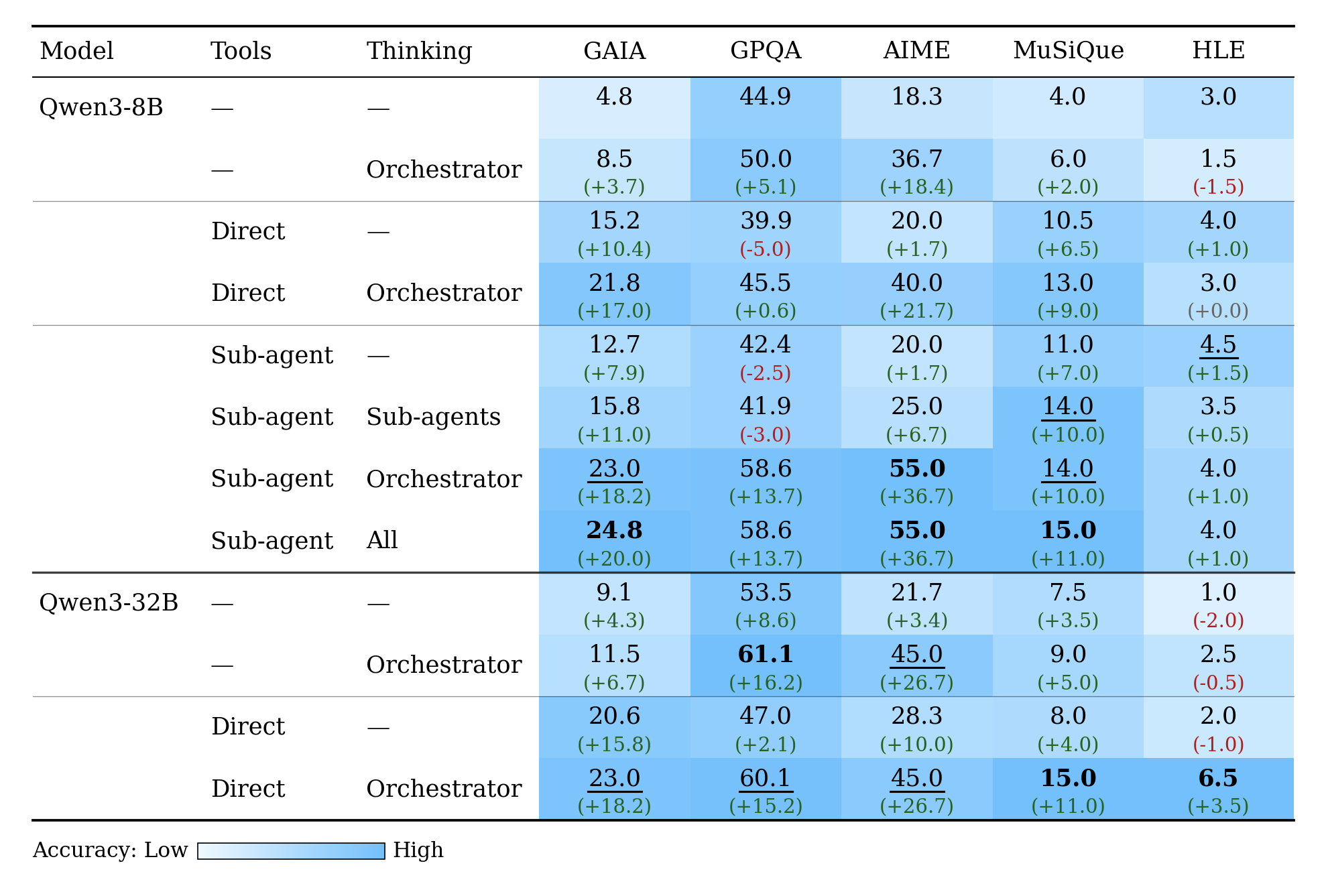}
    \label{tab:main_results}
\end{table}

To address RQ1, we compare the proposed multi-agent system against two strong single-agent baselines: (i) a larger 32B base model without tools and (ii) a 32B single-agent model equipped with direct tool use.

\textbf{A well-orchestrated 8B multi-agent system matches a 32B single-agent model with 
direct tools.} With orchestrator-only thinking, the 8B multi-agent system matches the 32B 
single-agent on GAIA (23.0 vs.\ 23.0) and substantially outperforms it on AIME (55.0 
vs.\ 45.0), falling slightly short only on GPQA (58.6 vs.\ 60.1) and MuSiQue (14.0 
vs.\ 15.0). Without thinking, the gap widens considerably, most notably on AIME (20.0 
vs.\ 28.3). HLE remains unsolved across all configurations, suggesting it exceeds the 
capabilities of current systems regardless of architecture.

\textbf{Tools help on retrieval tasks but hurt on knowledge-intensive reasoning; thinking 
helps broadly.} Within the 8B single-agent setting, direct tool use yields strong gains on 
GAIA (+10.4) and MuSiQue (+6.5), but degrades GPQA ($-5.0$), as indiscriminate retrieval 
overrides correct parametric knowledge. In contrast, orchestrator-only thinking produces 
consistent gains across all benchmarks (+3.7 GAIA, +5.1 GPQA, +18.4 AIME), and combining 
both yields the largest single-agent improvements overall (+17.0 GAIA, +21.7 AIME).

\textbf{Orchestrator thinking is the critical ingredient when transitioning to a 
multi-agent architecture.} Without thinking, moving to a multi-agent setup produces mixed 
results compared to single-agent direct tool use. Adding orchestrator-only thinking yields 
the strongest overall performance (+18.2 GAIA, +36.7 AIME, +10.0 MuSiQue), while enabling 
thinking for all agents provides only marginal further gains, indicating diminishing returns 
from distributing reasoning across every component. These findings motivate a deeper 
analysis of where thinking should be applied, which we explore in Section~\ref{sec:thinking_effect}.


\subsection{Effect of Explicit Thinking on Orchestrator and Sub-Agents} \label{sec:thinking_effect}

Given the performance gains observed from explicit thinking in the multi-agent setting, we further investigate how explicit thinking affects system performance and where it should be applied. In\,particular, we study (i) which components of the system benefit most from explicit thinking and (ii) when explicit thinking, and tool use, improves or degrades performance through qualitative analysis.



\textbf{Thinking at the orchestrator level drives performance gains; thinking in sub-agents 
provides little or no benefit.} Across all reasoning configurations (no thinking, 
sub-agents only, orchestrator only, all agents), orchestrator-only thinking yields the 
largest and most consistent gains (Table~\ref{tab:main_results}), while enabling thinking 
only for sub-agents leads to marginal improvements or degradation. This asymmetry reflects 
the distinct roles of each component: as the orchestrator governs global planning and task 
decomposition, explicit reasoning at this level directly improves system-wide decisions. 
Sub-agents, by contrast, execute localized subtasks, and thinking at this level introduces 
verbosity and coordination overhead without improving execution quality.

\textbf{Combining thinking with tools is most effective at the orchestrator level.} 
Explicit orchestrator reasoning enables more selective tool use: rather than dispatching 
tools indiscriminately, the orchestrator plans a targeted execution strategy before acting. 
This interaction explains why the combination of orchestrator thinking and sub-agent tools 
yields the strongest results, and why sub-agent thinking, which does not improve 
planning, fails to replicate these gains. We ablate individual MAS components in 
Appendix~\ref{app:ablation}; each addresses a distinct capability gap, with no single 
agent dominant across all benchmarks.

\subsection{Orchestrator Capacity vs.\ Sub-Agent Capacity}

\vspace{-1em}
\begin{figure}[h]
    \centering
    \includegraphics[width=0.8\textwidth]{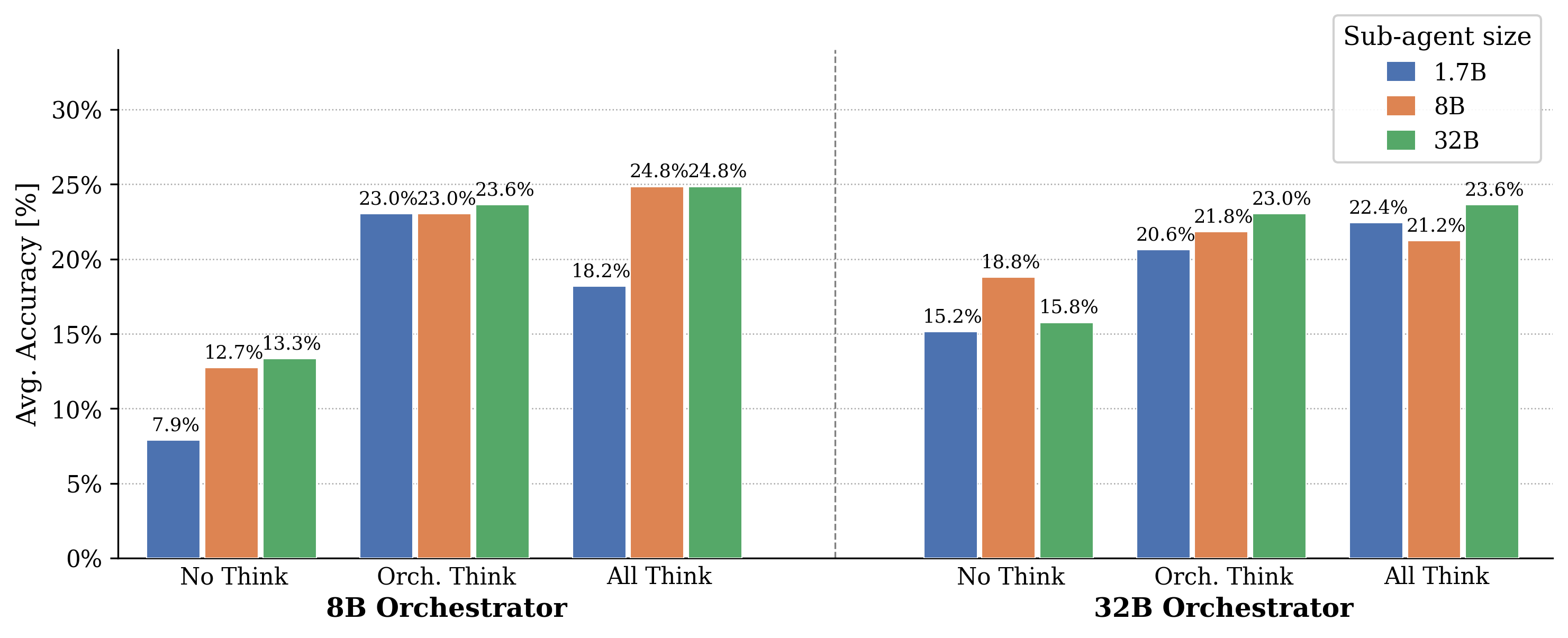}
    \caption{Comparison of 8B and 32B orchestrators paired with sub-agents ranging from 1.7B to 32B on GAIA. System performance is primarily driven by orchestrator thinking, while sub-agent scaling provides minimal and inconsistent gains.}
    \label{fig:orchestrator_vs_sub}
\end{figure}

As shown in Section~\ref{sec:main_results}, the largest gains from explicit thinking arise when applied at the orchestrator level. We now investigate how orchestrator capacity interacts with sub-agent size by systematically varying both (orchestrator: 8B vs.\ 32B; sub-agents: 1.7B, 8B, 32B).



\textbf{Once orchestrator thinking is enabled, sub-agent size becomes negligible.} 
Figure~\ref{fig:orchestrator_vs_sub} shows results on GAIA across all orchestrator and 
sub-agent size combinations. With orchestrator thinking enabled, the 8B orchestrator 
scores 23.0, 23.0, and 23.6 with 1.7B, 8B, and 32B sub-agents, respectively, effectively 
flat across sub-agent sizes. Without thinking, sub-agent size matters more (7.9 vs.\ 12.7 
vs.\ 13.3), confirming that a strong planning signal from the orchestrator compensates for 
weak executors.

\textbf{Scaling sub-agents is an inconsistent and inefficient use of compute.} Upgrading 
from an 8B to a 32B orchestrator with thinking yields only moderate gains, and is already 
less impactful than simply enabling thinking in the 8B orchestrator. Scaling sub-agents 
provides no consistent benefit and can hurt performance: under the 32B orchestrator with 
all thinking, 8B sub-agents (21.2) underperform both 1.7B (22.4) and 32B (23.6) 
counterparts. These results suggest that \textbf{multi-agent systems are planner-limited 
rather than executor-limited}: investment in orchestrator reasoning yields greater returns 
than investment in sub-agent scale.

\section{More Analysis}
\subsection{Efficiency--Performance Trade-offs}
\label{sec:efficiency_performance}
While small multi-agent systems achieve comparable accuracy to much larger single-agent models (Section~\ref{sec:main_results}), they involve more model calls due to sub-agent dispatch, raising the question of whether this coordination introduces more compute and latency costs. To quantify this, we measure mean wall-clock latency per query and total token consumption accumulated across all turns and sub-agent calls. All experiments run on 2$\times$H100 GPUs. Full details are in Appendix~\ref{app:efficiency}.


\begin{figure}[t]
    \centering
    \begin{minipage}{0.49\textwidth}
        \centering
        \includegraphics[width=\linewidth]{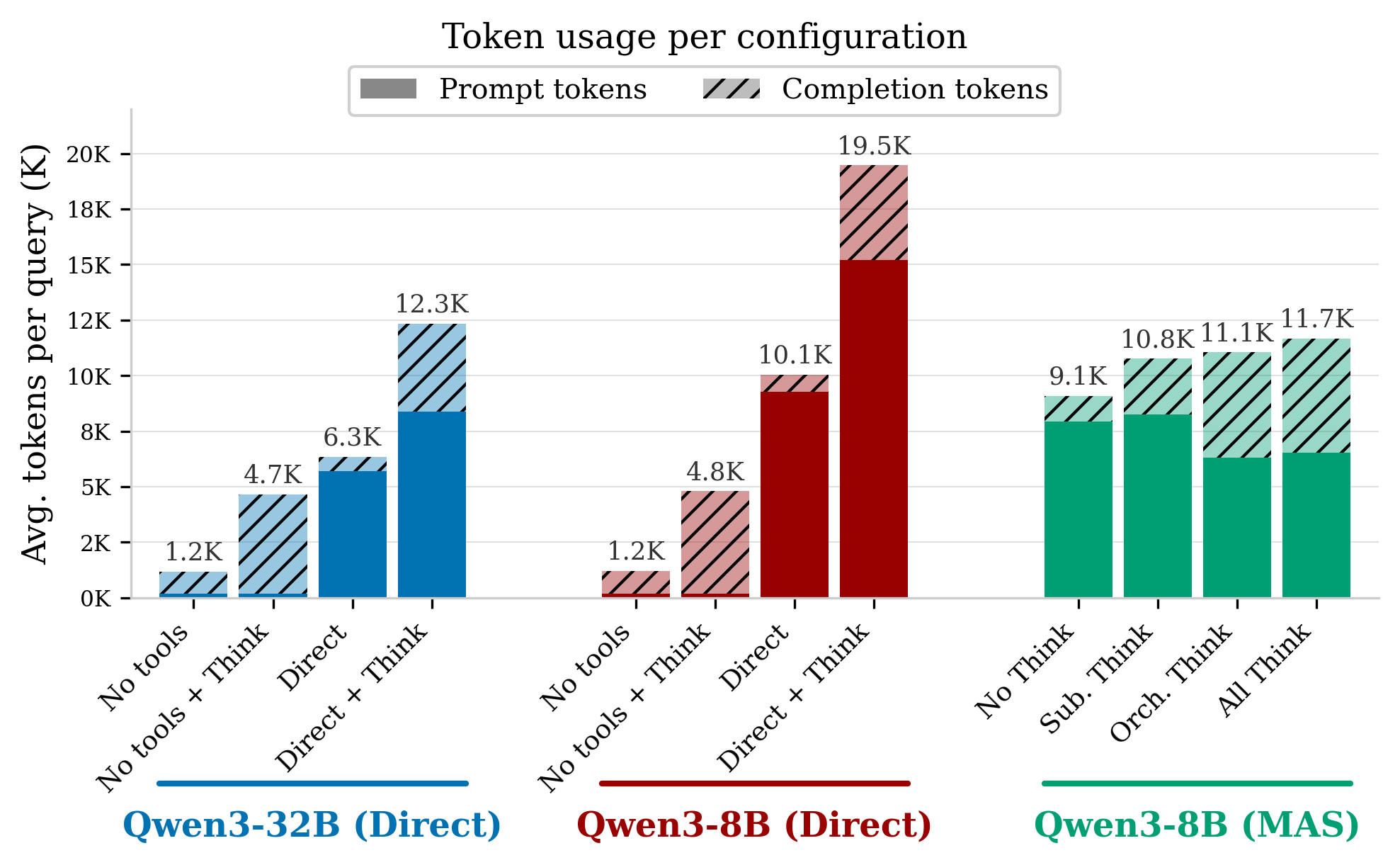}
    \end{minipage}\hfill
    \begin{minipage}{0.49\textwidth}
        \centering
        \includegraphics[width=\linewidth]{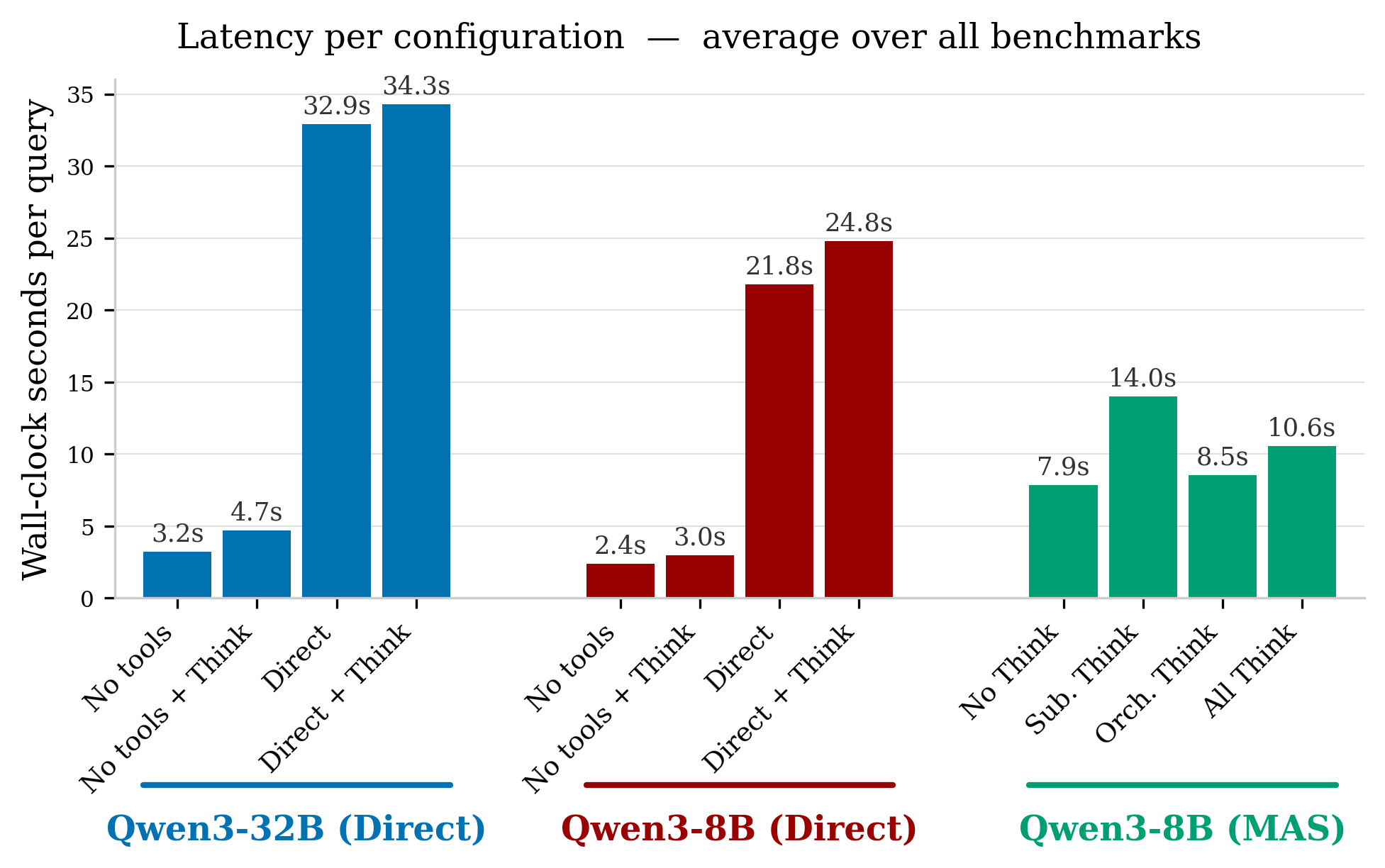}
    \end{minipage}
    \caption{Average token usage (left) and average latency per query (right) across configurations. Direct configurations accumulate prompt tokens across turns; MAS structured memory cuts this by 43\%. Thinking overhead is model-size-independent. Sub-agent thinking adds $+$77\% latency vs.\ $+$8\% for orchestrator thinking. 8B MAS is $4.2\times$ faster than 32B direct despite more model calls.}
    \label{fig:efficiency}
\end{figure}

\paragraph{Structured memory contains token growth.}
Single-agent direct runs re-feed the full conversation history each turn,
so prompt tokens grow with every tool call, adding thinking nearly doubles
usage from 10.1K to 19.5K for the 8B model.
MAS builds a compact summary at each turn instead, holding usage
to 9.1--11.7K across \emph{all} thinking modes, a 43\% reduction over
\textit{Direct\,+\,Think}.

\paragraph{Thinking overhead depends on where it is applied.}
Orchestrator thinking reduces turns (3.6$\to$2.0) and tool calls
(2.7$\to$1.1): the planning trace drives more decisive action, so saved
steps offset longer generation, adding only $+$0.6\,s.
Sub-agent thinking leaves turn count unchanged (3.6$\to$3.8) but attaches a
reasoning trace to each of the ${\approx}$2.9 sequential tool calls,
multiplying decoding cost and adding 6.1\,s.

\paragraph{MAS is faster than single-agent direct despite more model calls.}
The 8B MAS without thinking (7.9\,s) is $2.8\times$ faster than 8B direct
(21.8\,s) and $4.2\times$ faster than 32B direct (32.9\,s), as compact
per-turn prompts, short single-turn sub-agent calls, and faster 8B inference
together more than offset the cost of multiple serialised generations.

\subsection{Tool Usage Patterns}

\begin{figure}[H]
    \centering
    \includegraphics[width=\textwidth]{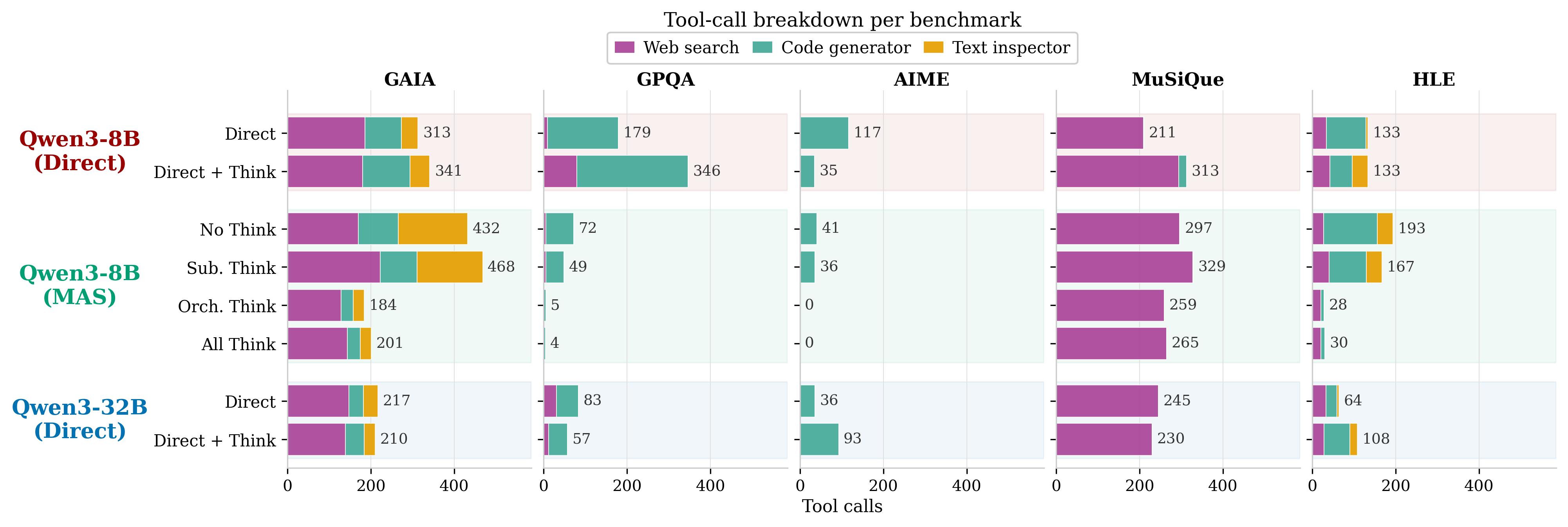}
    \caption{Tool call distributions across benchmarks and configurations. Orchestrator thinking shifts the distribution toward targeted, task-appropriate tool calls compared to no-thinking baselines.}
    \label{fig:tool_calls_breakdown}
\end{figure}

To better understand behavioral differences between single-agent, MAS and the role of explicit thinking, we analyze the distribution and volume of tool calls across different settings. Figure~\ref{fig:tool_calls_breakdown} summarizes tool usage patterns for 8B and 32B models under direct and multi-agent configurations.

\vspace{-1em}

\paragraph{Orchestrator thinking halves tool call volume.}
The most striking pattern is how orchestrator thinking reduces total tool calls for the 8B MAS, nearly halving execution volume. In contrast, sub-agent-only thinking is counterproductive: it slightly \emph{increases} calls on GAIA (447$\to$483) while degrading performance, consistent with the finding that sub-agent reasoning traces introduce coordination overhead without improving global planning.

\paragraph{Benchmark composition drives tool usage patterns.}
AIME tool calls collapse to zero under orchestrator thinking, as the model solves problems through its reasoning trace rather than dispatching tools, appropriate for symbolic reasoning tasks. GPQA shows a similar drop (72$\to$5). MuSiQue remains tool-heavy across all configurations ($\sim$259--329), reflecting that multi-hop retrieval genuinely requires web search regardless of reasoning mode.

\paragraph{The 8B MAS makes fewer but better tool calls than a 32B single agent.}
The 32B single-agent with orchestrator thinking uses 698 total tool calls, yet the 8B MAS with orchestrator thinking reaches only 476--32\% fewer calls while achieving competitive accuracy which suggests that effective orchestration produces more selective tool use than raw model scale alone.

\subsection{Qualitative Case Analysis}
We qualitatively analyze when tools and thinking improve or degrade performance. Full case studies are in Appendix~\ref{app:qualitative_analysis_tools}.
\paragraph{When tools help.} Tool access helps most when the task requires information the model cannot supply from parameters alone. Multi-hop questions benefit because iterative search calls ground each intermediate entity against retrieved evidence, preventing error propagation across hops. Similarly, tasks that attach structured files are unsolvable without tools: a text-inspector sub-agent must first extract the content before a code sub-agent can operate on it. Even single-hop questions gain when the critical fact is sufficiently obscure, as chained search calls can locate details that parametric memory lacks.
\paragraph{When tools hurt.} Tools degrade performance when retrieved content overrides correct parametric knowledge, for instance, a wrong identifier propagated through downstream computation, or when search returns nothing and the model defaults to ``insufficient information'' rather than falling back on a correct memory-based answer. In both cases the model over-trusts the tool signal and discounts what it already knows.
\paragraph{When thinking helps.} Without explicit thinking, the orchestrator suffers from premature guessing and misdirected tool calls. Thinking addresses this along three axes: task decomposition (planning multi-step retrieval rather than halting at the first plausible guess), tool selection (detecting unavailable tools and rerouting, or chaining multiple tools with a pre-planned computation schema), and constraint preservation (encoding disambiguating constraints into queries and maintaining adherence to complex instructions that surface-level reading would misinterpret).
\paragraph{When thinking hurts.} Thinking introduces regressions when reasoning and action become misaligned. The orchestrator may solve a problem entirely within its trace without dispatching the appropriate tool, bias a search query toward confirming a strong prior rather than retrieving the answer, or conclude that a resource is inaccessible without ever attempting retrieval. The common thread is over-deliberation: thinking substitutes internal rationalization for external verification.



\section{Conclusion}
This work investigated whether collaborative systems of smaller language models can rival larger monolithic models and analyzed the roles of orchestration, model scale, and explicit reasoning. On the GAIA benchmark, we find that well-orchestrated small models can match or exceed larger single-agent baselines, with performance driven primarily by the capacity of the Orchestrator rather than the size of execution sub-agents. We also identify trade-offs in explicit reasoning: reasoning at the Orchestrator level improves task decomposition and completion, whereas excessive reasoning can introduce inefficiencies such as tool-call thrashing and output drift. Overall, these results suggest that effective system design, particularly the separation of planning and execution, can be as\,important as model scale in agentic settings.

\section*{Limitations}
We focus on the Qwen model family to enable controlled comparisons across scales. It remains to be seen whether the same trends hold for other model families with different architectures or training regimes. Our evaluation centers on reasoning- and knowledge-intensive benchmarks, including tool use, scientific QA, mathematics, multi-hop QA, and expert knowledge. While these tasks capture structured reasoning, they do not cover the full range of language model capabilities, such as open-ended or creative generation. In addition, we study a minimal multi-agent setup with a single orchestrator and a small set of specialized sub-agents under restricted communication. This design isolates the role of orchestration, but does not explore more flexible coordination or communication strategies that could further influence performance.

\section*{Acknowledgments}
We thank the members of the IRLab at the University of Amsterdam and the LESSEN project for their valuable feedback on our experiments. This research was partially supported by the Dutch Research Council (NWO) under project numbers 024.004.022, NWA.1389.20.183, and KICH3.LTP.20.006, and by the European Union under grant agreement No. 101201510 (UNITE). The views and opinions expressed are those of the authors only and do not necessarily reflect those of their respective employers, funders, or granting authorities.

\bibliography{iclr2026_malgai}
\bibliographystyle{iclr2026_conference}

\appendix
\section{Benchmarks} \label{app:benchmarks}
We evaluate on five benchmarks that collectively span the core capabilities required by agentic and reasoning systems.

\textbf{GAIA} \citep{mialon2023gaiabenchmarkgeneralai} is designed to 
evaluate agentic behavior through tool-intensive tasks. It comprises three 
difficulty levels: Level~1 tests factual retrieval and shallow reasoning, 
Level~2 emphasizes multi-hop reasoning and planning, and Level~3 requires 
complex coordination between reasoning and tools. We use the validation 
split of 165 questions. 

\textbf{GPQA} \citep{rein2023gpqagraduatelevelgoogleproofqa} 
(Graduate-Level Google-Proof Q\&A) is a multiple-choice benchmark of 
challenging questions in biology, chemistry, and physics, written by domain 
experts and designed to be difficult even for highly skilled non-experts 
with web access. We evaluate on the Diamond split (198 questions).

\textbf{AIME} \citep{aime} consists of problems from the American Invitational 
Mathematics Examination, requiring precise multi-step algebraic and 
combinatorial reasoning. We evaluate on the 2024 and 2025 problem sets, 
comprising 60 problems in total.

\textbf{MuSiQue} \citep{trivedi2022musiquemultihopquestionssinglehop} is 
a multi-hop question answering dataset that requires composing information 
across multiple documents, testing systematic retrieval and reasoning 
chains. We evaluate on a randomly sampled subset of 200 questions from the 
validation set.

\textbf{HLE} \citep[Humanity's Last Exam,][]{phan2025lastexam} is a 
recently introduced benchmark of difficult questions across 
diverse academic domains, designed to be at the frontier of human expert 
knowledge. We evaluate on a randomly sampled subset of 200 questions from 
the test set.

\section{Ablation Study}
\label{app:ablation}

Table~\ref{tab:unified_ablation} shows the effect of ablating each MAS component individually, highlighting their contributions to task performance.
The Web Searcher is the most critical for retrieval-heavy tasks, reducing GAIA by 9.7~pp and MuSiQue by 10~pp when removed.  
The File Inspector is important for GAIA (\(-6.6\)~pp), where some tasks involve file attachments.  
Structured Memory has the largest impact on multi-step reasoning: removing it collapses AIME by 18.3~pp and GPQA by 8.1~pp, as accumulated context is essential for long reasoning chains.  
In contrast, removing the Coder slightly reduces GAIA performance, but marginally improves AIME and GPQA.
Overall, these results indicate that each component addresses a distinct capability gap, and no single agent is universally dominant across all task types.

\begin{table}[H]
\centering
\caption{Ablations across tools and structured memory (Qwen3-8B, sub-agent tools,
orchestrator-only thinking). Each row removes one component from the full system.
A dash (--) indicates the component was not applicable or not used for that dataset.
Bold denotes the best result per dataset; underline denotes the second best.
}
\label{tab:unified_ablation}
\vspace{0.5em}
\begin{tabular}{ccccrrrrr}
\toprule
\textbf{Web} & \multirow{2}{*}{\textbf{Coder}} & \textbf{File} & \textbf{Structured} & \multirow{2}{*}{\textbf{GAIA}} & \multirow{2}{*}{\textbf{AIME}} & \multirow{2}{*}{\textbf{GPQA}} & \multirow{2}{*}{\textbf{HLE}} & \multirow{2}{*}{\textbf{MuSiQue}} \\
\textbf{searcher} &  & \textbf{inspector} & \textbf{memory} &  &  &  &  &  \\
\midrule
\cmark & \cmark & \cmark & \cmark & \textbf{23.0} & 55.0 & \underline{58.6} & \textbf{4.0} & 14.0 \\
\xmark & \cmark & \cmark & \cmark & 13.3 & \underline{58.3} & 57.1 & \underline{3.0} & 4.0 \\
\cmark & \xmark & \cmark & \cmark & 21.2 & \textbf{60.0} & \textbf{59.6} & \underline{3.0} & \underline{15.5} \\
\cmark & \cmark & \xmark & \cmark & 16.4 & -- & -- & 2.0 & -- \\
\cmark & \cmark & \cmark & \xmark & \underline{21.8} & 36.7 & 50.5 & 2.5 & \textbf{18.0} \\
\bottomrule
\end{tabular}
\end{table}
\section{Efficiency: Token Usage and Latency Details}
\label{app:efficiency}

\subsection{Token Counting}

Token counts represent the \textbf{total accumulated prompt and completion
tokens} across all turns of a single query, including all sub-agent calls.
Prompt tokens are measured as the tokenizer-encoded length of the rendered
input at each model call; completion tokens
are the length of the decoded output.
Both are summed via a running accumulator that is updated after every
model generation, including the planning turn and all sub-agent dispatches.
Latency is the mean wall-clock time per query,
$(t_\text{end} - t_\text{start}) / N$, measured end-to-end for each run
on 2$\times$H100 GPUs.

\subsection{How Prompt Tokens Accumulate}

\paragraph{Direct (baseline) configurations.}
In single-agent direct runs the model receives the full conversation history
at every turn.
The prompt for turn $n$ consists of the original system and user messages
followed by all prior assistant responses and tool outputs:
\[
  \text{prompt}_n = [\,\text{sys},\;\text{user},\;a_1,\;t_1,\;\ldots,\;a_{n-1},\;t_{n-1}\,],
\]
where $a_i$ and $t_i$ are the assistant response and tool output at step $i$.
Prompt size therefore grows linearly with the number of tool calls, and the
cumulative total across all turns grows quadratically, with prompt tokens
dominating the total in multi-turn interactions.

\paragraph{MAS structured-memory configurations.}
AgentFlow's orchestrator does not replay the full message history.
At each turn it builds a fresh two-message prompt:
\begin{itemize}[leftmargin=*,nosep]
  \item \textbf{System:} the orchestrator's role-specific system prompt.
  \item \textbf{User:} the original query, a planning trace generated at
        turn~0 (\emph{query analysis}), and a compact per-step record of
        completed actions (\emph{action history}: tool name, sub-goal, and a
        brief result summary).
\end{itemize}
This summary replaces the raw message replay: each step
contributes a fixed-length record rather than the full assistant response and raw tool output, so prompt size grows slowly with step count.

\paragraph{Sub-agent calls.}
Each sub-agent dispatch (web search, code generation, text inspection) is a
\emph{single-turn} call: the sub-agent receives its own system prompt and a
concise task description, produces one response, and returns.
Its tokens are counted and added to the running total for the query.
When sub-agent thinking is enabled, each dispatch also generates a reasoning
trace before the response that is passed back to the orchestrator, increasing completion tokens per call.

\subsection{Why Orchestrator and Sub-Agent Thinking Have Asymmetric Cost}
Orchestrator thinking prepends a planning trace before each action, which
leads to more decisive execution: turns and tool calls roughly halve relative
to the no-thinking baseline, as better upfront planning eliminates corrective
follow-up steps. The latency saved by fewer dispatches largely offsets the
cost of generating the trace, yielding a small net increase of only $+$0.6\,s. Sub-agent thinking, by contrast, does not reduce the number of dispatches, 
the orchestrator still issues a similar number of tool calls, but attaches a reasoning trace to each one, which increases latency by 6.1\,s.

\subsection{Per-dataset Latency}

Table~\ref{tab:latency_heatmap} shows mean wall-clock latency per query
broken down by dataset and configuration.

\begin{table}[H]
    \centering
    \caption{Per-dataset mean latency (seconds per query) across all
             configurations. MAS configurations are consistently faster
             than single-agent direct runs on tool-intensive datasets,
             while no-tool baselines are uniformly fast across all
             benchmarks.}
    \label{tab:latency_heatmap}
    \includegraphics[width=\textwidth]{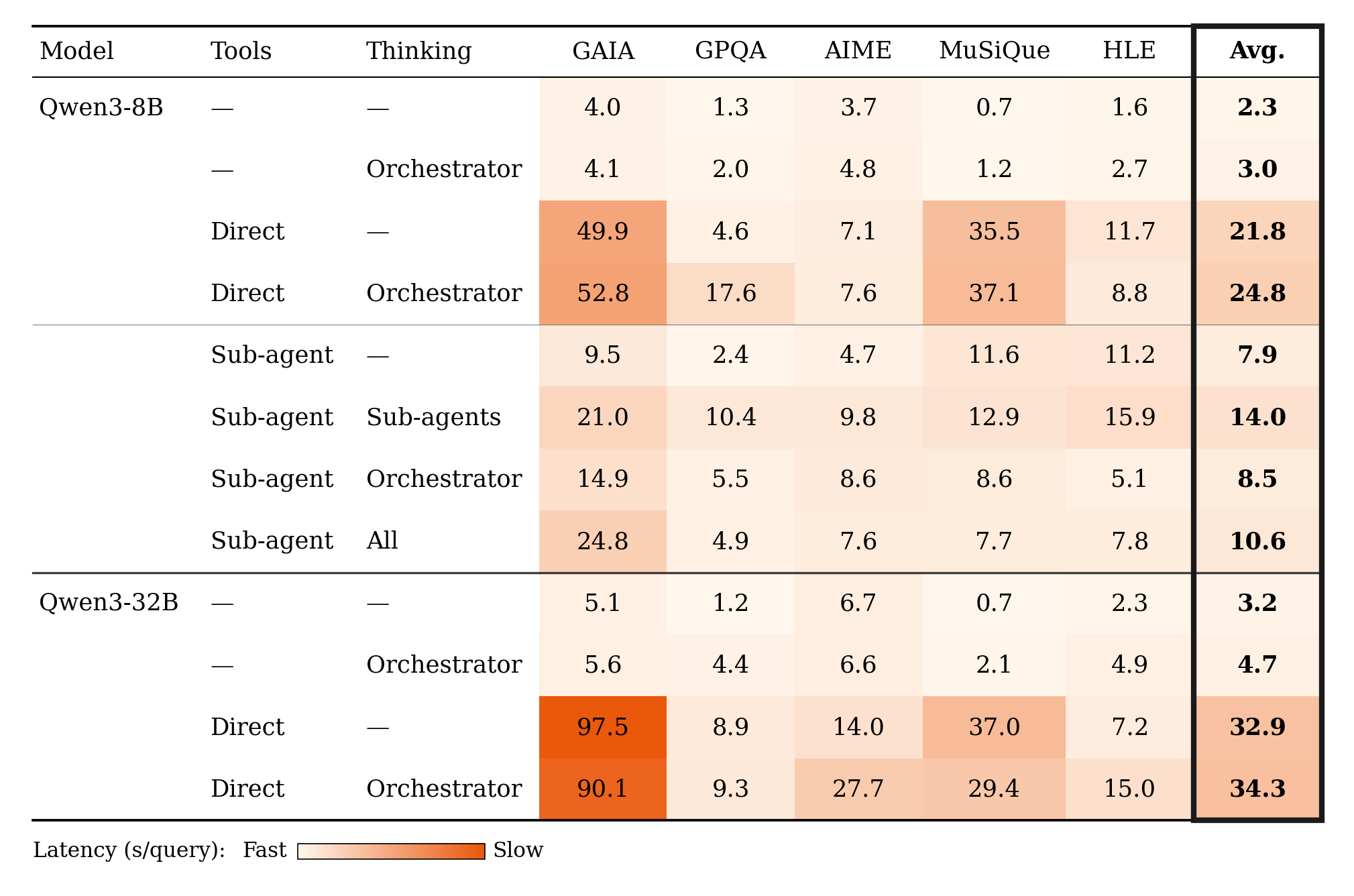}
\end{table}

\definecolor{qaframe}{RGB}{45, 95, 170}
\definecolor{qahead}{RGB}{240, 245, 255}
\definecolor{qapassfg}{RGB}{30, 100, 50}
\definecolor{qafailfg}{RGB}{160, 50, 40}
\definecolor{qameta}{RGB}{100, 110, 130}

\makeatletter
\newcommand{\qa@row}[5]{%
  \begingroup
  \setlength{\fboxsep}{0pt}%
  \noindent
  \begin{minipage}{\linewidth}%
      \vspace{2pt}
      \hspace{4pt}{\scriptsize\sffamily\bfseries\color{#2}#1\ #3}\par
      \vspace{0pt}
      \hspace{4pt}{\scriptsize\sffamily\color{black}#4}\par
      \vspace{-5pt}
      \hspace{4pt}\parbox[t]{\dimexpr\linewidth-4pt\relax}{%
        \tiny\sffamily\color{qameta}#5%
      }\vspace{2pt}
  \end{minipage}%
  \endgroup
}

\newcommand{\qafail}[3]{%
  \qa@row{#1}{qafailfg}{\xmark}{#2}{#3}%
  \vspace{1pt}\par\color{qaframe}\hrule height 0.4pt\vspace{2pt}
}

\newcommand{\qapass}[3]{%
  \qa@row{#1}{qapassfg}{\cmark}{#2}{#3}%
  \vspace{1pt}\par\color{qaframe} \hrule height 0.4pt\vspace{2pt}
}
\makeatother

\newenvironment{qablock}[2]{%
  \leavevmode\par
  \begin{tcolorbox}[
    enhanced,
    colback=white, colframe=qaframe,
    boxrule=0.6pt, arc=3pt,
    left=1pt, right=1pt, top=0pt, bottom=0pt,
    toptitle=2pt, bottomtitle=2pt,
    colbacktitle=qahead, coltitle=black,
    fonttitle=\tiny\sffamily\color{qameta},
    title={#1\\[1pt] {\scriptsize\bfseries #2}},
    before upper={\vspace{0pt}},
    after upper={\vspace{0pt}}
  ]%
}{%
  \end{tcolorbox}\vspace{1pt}%
}

\newcommand{\mockqablock}{
\begin{qablock}{Qwen3-8B}{Example Question}
\qafail{Setup 1 (fail)}{Answer Text}{Comment on tools}
\qapass{Setup 2 (pass)}{Answer Text}{Comment on tools}
\end{qablock}
}


\section{Qualitative Analysis of Tool Usage and Thinking}
\label{app:qualitative_analysis_tools}

This appendix presents representative examples from four benchmarks (GAIA, MuSiQue,
AIME, GPQA) that illustrate when tool access helps or hurts, and when explicit
orchestrator thinking helps or hurts. We use three configurations:
\emph{No Tools} - no tool access, no thinking;
\emph{Tools} - sub-agent tools available, no orchestrator thinking;
\emph{Tools + Thinking} - sub-agent tools plus orchestrator thinking (our primary setup).
Sections~\ref{app:tools_help} and~\ref{app:tools_hurt} compare
\emph{No Tools} vs.\ \emph{Tools} to isolate the effect of tool access.
Sections~\ref{app:thinking_helps}–\ref{app:thinking_hurts} then compare
\emph{Tools} vs.\ \emph{Tools + Thinking} to isolate the effect of explicit reasoning.
All examples use Qwen3-8B for both Orchestrator and sub-agents.

\subsection{When tools improve performance}
\label{app:tools_help}

The clearest gains from tool access occur on tasks that require grounding
multi-hop chains of inference in retrieved evidence, or that require inspecting
file attachments the model cannot read directly.

\paragraph{Multi-hop web retrieval.}
Multi-hop questions need each intermediate entity resolved against external
evidence before the final answer can be reached. Without tools, the model
guesses intermediate facts from memory and propagates confabulated errors.
With web search, repeated targeted queries
iteratively ground each hop.

\begin{qablock}{Qwen3-8B}%
  {Name the castle in the city where the performer of \emph{Never Too Loud}
   was formed. \textnormal{\tiny(MuSiQue)}}
\qafail{No Tools}{Tower of London}%
  {No tool calls. Confabulates the performer's home city as London; finds
   the wrong castle.}
\qapass{Tools}{Casa Loma}%
  {Tool calls: web\_search $\times$5. After three empty results, a refined
   query identifies the band as Danko Jones (Toronto); a follow-up search
   returns Casa Loma as the castle in Toronto.}
\end{qablock}

\paragraph{File inspection + code execution.}
Some tasks attach structured files that cannot be read from parametric
memory. A text-inspector sub-agent reads the file; a code sub-agent executes
graph algorithms on the extracted data. Without tools the model cannot
even access the input.

\begin{qablock}{Qwen3-8B}%
  {Each cell in the attached spreadsheet represents a land plot (green =
   eligible). Is a Hamiltonian circuit possible? \textnormal{\tiny(GAIA)}}
\qafail{No Tools}{Cannot determine without the actual data}%
  {No tool calls. Model acknowledges it cannot access the attachment and
   returns no answer.}
\qapass{Tools}{No}%
  {Tool calls: text\_inspector $\times$1 (reads colour grid),
   code\_generator $\times$2 (parses adjacency, runs Hamiltonian-circuit
   check). Returns the correct answer programmatically.}
\end{qablock}

\paragraph{Two-step factual disambiguation.}
Some questions require retrieving an obscure factual link that is absent
from parametric memory. A sequence of web-search calls, each building on
the previous result, locates the right detail.

\begin{qablock}{Qwen3-8B}%
  {An AI-regulation arXiv paper (Jun.\ 2022) shows a 3-axis figure. Which
   axis label also names a society type in a 2016 Physics \& Society paper?
   \textnormal{\tiny(GAIA)}}
\qafail{No Tools}{Autonomy}%
  {No tool calls. Guesses from parametric knowledge; wrong.}
\qapass{Tools}{egalitarian}%
  {Tool calls: web\_search $\times$2. First search finds the 2016 Physics
   \& Society paper whose axis label ``egalitarian'' matches. Second
   confirms the link.}
\end{qablock}

\subsection{When Tools Degrade Performance}
\label{app:tools_hurt}

Tools can actively harm performance when retrieved content contradicts correct parametric knowledge, or when a tool returns nothing and the model abandons a correct memory-based answer in favour of ``insufficient information.''

\paragraph{Search-returned wrong identifier.}
The model's parametric knowledge encodes the correct value of a specific
identifier. The web-search sub-agent retrieves a different (incorrect) value,
which the code sub-agent then uses to compute the wrong final result.

\begin{qablock}{Qwen3-8B}%
  {Compute the ISBN-10 check digit of the Tropicos ID for
   \emph{Helotiales}. \textnormal{\tiny(GAIA)}}
\qapass{No Tools}{3}%
  {No tool calls. Computes directly from the correct Tropicos ID stored
   in parametric memory; answer is right.}
\qafail{Tools}{8}%
  {Tool calls: web\_search $\times$1 (returns wrong ID 100370510),
   code\_generator $\times$1 (computes check digit for the wrong number).
   Correct parametric knowledge is overwritten by a faulty retrieval.}
\end{qablock}

\paragraph{Search blind spot on rare facts.}
For questions that require obscure historical knowledge, web search may
return no useful content. When all queries fail, the model
returns an empty answer, whereas the model without tools might still produce the correct answer from parametric knowledge.

\begin{qablock}{Qwen3-8B}%
  {Which city, linked to a sea Theodoric was trying to access by attacking
   Narbo Martius, became the capital of Ethiopia?
   \textnormal{\tiny(MuSiQue)}}
\qapass{No Tools}{Adulis}%
  {No tool calls. Retrieves from parametric historical knowledge; correct.}
\qafail{Tools}{(empty)}%
  {Tool calls: web\_search $\times$2. Both queries return ``No helpful
   information found.'' Orchestrator gives up and returns an empty
   prediction rather than falling back to parametric knowledge.}
\end{qablock}

\subsection{When Thinking Improves Tool Use}
\label{app:thinking_helps}

In these cases, the \emph{Tools} setting (no orchestrator thinking) fails
due to premature guessing or misdirected tool use. Explicit thinking bridges
this gap by enforcing task decomposition, maintaining adherence to constraints,
and selecting the most effective tool for each sub-problem.

\paragraph{Task decomposition (verification).}
Without thinking, the orchestrator often halts after generating a plausible
guess from parametric memory, never verifying it with external sources. Thinking
recognises that memory alone is insufficient, plans a multi-step retrieval
pipeline, and executes it to produce a verified answer.

\begin{qablock}{Qwen3-8B}%
  {Of the cities within the U.S.\ where presidents were born, which two
   are the farthest apart geographically? Give them alphabetically.
   \textnormal{\tiny(GAIA)}}
\qafail{Tools}{Charleston,\ Hartford}%
  {No tool calls. Guesses two plausible-sounding cities from parametric
   memory; both are wrong.}
\qapass{Tools + Thinking}{Braintree,\ Honolulu}%
  {Tool calls: web\_search $\times$2, code\_generator $\times$1.
   Thinking reasons: ``I can't recall all 46 birth cities; I need to
   retrieve the full list, then compute the east--west span.'' Two
   searches build the complete list; code computes the geographic
   extremes; result is verified and sorted.}
\end{qablock}

\paragraph{Accurate tool selection.}
Without thinking, the orchestrator dispatches a non-existent tool
(\texttt{video\_analysis} is not available in the system) for a video-URL
task and retries it blindly until the turn limit is exhausted. Thinking
detects that the tool does not exist, identifies an alternative retrieval
route, and redirects in a single call.

\begin{qablock}{Qwen3-8B}%
  {What does Teal'c say in response to ``Isn't that hot?'' in
   \texttt{youtu.be/1htKBjuUWec}? \textnormal{\tiny(GAIA)}}
\qafail{Tools}{(empty - 15 turns exhausted)}%
  {Tool calls: video\_analysis $\times$15. Tool does not exist in the
   system; every call returns empty. Without thinking to detect this,
   the orchestrator keeps retrying until the turn limit is reached.}
\qapass{Tools + Thinking}{Extremely}%
  {Tool calls: web\_search $\times$1; single query retrieves the
   exact line.}
\end{qablock}

\paragraph{Accurate tool Selection (chaining).}
Thinking enables the orchestrator to combine tools when a
single tool is insufficient. Here, reading the file is necessary but not
enough: thinking plans the computation schema \emph{before} the tool is
called, so the returned data is immediately actionable.

\begin{qablock}{Qwen3-8B}%
  {Attached file: vendors in Liminal Springs mall with monthly revenue
   and rent. Which vendor has the lowest revenue-to-rent ratio? Return
   its ``type'' field. \textnormal{\tiny(GAIA)}}
\qafail{Tools}{(empty)}%
  {Tool calls: text\_inspector $\times$2. File is read correctly, but
   without a computation plan the orchestrator cannot synthesise
   ``revenue $\div$ rent per row'' from raw table output.}
\qapass{Tools + Thinking}{Finance}%
  {Tool calls: text\_inspector $\times$2. Thinking first plans:
   ``compute revenue$\div$rent for each vendor, find the minimum,
   return its type field.'' File data is then correctly reduced;
   orchestrator extracts \emph{Finance}.}
\end{qablock}

\paragraph{Constraint preservation (disambiguation).}
When a question requires identifying a specific entity among many similarly
named ones, the no-thinking orchestrator latches onto the first plausible
match. Thinking generates a more precise search query that encodes the
distinguishing constraint, bypassing the common distractor.

\begin{qablock}{Qwen3-8B}%
  {What two-word model type did the 2018--19 studies by Manash Pratim
   Kashyap and PS Fader on customer retention have in common?
   \textnormal{\tiny(GAIA)}}
\qafail{Tools}{discrete time}%
  {Tool calls: web\_search $\times$1. Generic query returns a broad
   churn-modelling overview; orchestrator selects the first plausible
   model type.}
\qapass{Tools + Thinking}{beta geometric}%
  {Tool calls: web\_search $\times$1. Thinking recalls that PS Fader is
   specifically associated with probabilistic customer-lifetime models,
   generates the targeted query ``Kashyap Fader customer retention
   common model type 2018 2019'', and retrieves the specific
   \emph{beta geometric} identification.}
\end{qablock}

\paragraph{Constraint preservation (instruction adherence).}
Thinking helps the model maintain strict adherence to multi-layered
instruction traps, even before any tool is used. Here, the question is
encoded in reverse; the no-thinking orchestrator misreads the surface
text, while thinking first decodes it and then follows the instruction.

\begin{qablock}{Qwen3-8B}%
  {\texttt{.rewsna eht sa "tfel" drow eht fo etisoppo eht etirw
   ,ecnetnes siht dnatsrednu uoy fI}
   \textnormal{\tiny(GAIA)}}
\qafail{Tools}{123}%
  {No tool calls. Misreads the reversed text as a numeric sequence;
   outputs a nonsensical answer.}
\qapass{Tools + Thinking}{right}%
  {Tool calls: code\_generator $\times$1. Thinking recognises the
   string is reversed; dispatches code to reverse it, recovering
   the instruction ``write the opposite of the word `left'.'' Then
   follows the instruction: \emph{right}.}
\end{qablock}

\subsection{When Thinking Helps (Without Tool Use)}
\label{app:thinking_helps_notool}

On reasoning-intensive benchmarks, the dominant benefit of orchestrator
thinking is not tool coordination but pure inference quality. In these
examples neither configuration makes any tool call; the only difference
is whether an explicit reasoning trace is produced. Thinking activates
domain knowledge, applies mathematical identities, and
cross-checks intermediate steps before committing to an answer.

\paragraph{Domain-knowledge activation (science).}
Expert multiple-choice questions require applying a non-obvious
mathematical identity before the correct answer class becomes apparent.
Without thinking, the model applies the superficially similar isotropic
formula. With thinking, it decomposes the potential using
$\cos^2\theta = \tfrac{1+\cos 2\theta}{2}$, reveals the anisotropic
oscillator structure, and selects the correct energy spectrum.

\begin{qablock}{Qwen3-8B}%
  {$V(r,\theta)=\tfrac{1}{2}kr^2+\tfrac{3}{2}kr^2\cos^2\theta$.
   Find the energy spectrum.
   \textnormal{\tiny(GPQA)}\\[2pt]
   \textnormal{\tiny
     (A)~$E=(3n_x+2n_y+\tfrac{1}{2})\hbar\sqrt{k/m}$\quad
     (B)~$E=(n_x+3n_y+\tfrac{3}{2})\hbar\sqrt{k/m}$\\
     (C)~$E=(2n_x+3n_y+\tfrac{1}{2})\hbar\sqrt{k/m}$\quad
     (D)~$E=(2n_x+n_y+\tfrac{3}{2})\hbar\sqrt{k/m}$}}
\qafail{Tools}{C}%
  {No tool calls. Treats potential as isotropic 2D harmonic oscillator;
   wrong frequency coefficients.}
\qapass{Tools + Thinking}{D}%
  {No tool calls. Thinking applies $\cos^2\theta = \tfrac{1+\cos 2\theta}{2}$,
   identifies $\omega_x{=}\sqrt{k/m}$, $\omega_y{=}2\sqrt{k/m}$, and
   selects the correct anisotropic energy formula.}
\end{qablock}

\paragraph{Step-by-step chained arithmetic.}
Multi-step dex-notation problems require chaining several ratio
conversions. Without thinking, the model sketches the
correct formula but evaluates it incorrectly. With thinking, each
conversion is performed explicitly and verified before the next step.

\begin{qablock}{Qwen3-8B}%
  {$[\text{Si/Fe}]_1{=}0.3$, $[\text{Mg/Si}]_2{=}0.3$, solar
   abundances given. Find $n_{\text{Si},1}/n_{\text{Si},2}$.
   \textnormal{\tiny(GPQA)}\\[2pt]
   \textnormal{\tiny
     (A)~${\approx}1.2$\quad
     (B)~${\approx}0.8$\quad
     (C)~${\approx}12.6$\quad
     (D)~${\approx}3.9$}}
\qafail{Tools}{D}%
  {No tool calls. Correct formula identified but one dex step is
   evaluated incorrectly; selects D.}
\qapass{Tools + Thinking}{C}%
  {No tool calls. Thinking converts each $[\text{X/Y}]$ value
   step-by-step to a linear ratio, chains $n_{\text{Si},1}/n_H$ and
   $n_{\text{Si},2}/n_H$ separately, and correctly computes
   $\approx 12.6$.}
\end{qablock}

\subsection{When Thinking Hurts}
\label{app:thinking_hurts}

In these instances, explicit thinking induces regressions due to a
misalignment between reasoning and action. The orchestrator may
erroneously rationalise skipping tools, corrupt an otherwise effective
search query, or reason itself into inaction before attempting any
retrieval.

\paragraph{Skipping required tools.}
Thinking can lead the orchestrator to attempt to solve a computational
problem entirely within its reasoning trace, concluding with a confident
(but wrong) answer and never dispatching the code tool that the
no-thinking baseline correctly uses.

\begin{qablock}{Qwen3-8B}%
  {$N$ = \#\{8-digit permutations of $1\ldots8$ divisible by~22\}.
   Find $|N-2025|$. \textnormal{\tiny(AIME)}}
\qapass{Tools}{279}%
  {Tool calls: code\_generator $\times$1. Enumerates all 40\,320
   permutations, filters by divisibility~22, returns $N{=}1746$,
   $|1746{-}2025|{=}279$. Exact.}
\qafail{Tools + Thinking}{2583}%
  {No tool calls. Thinking applies alternating-digit-sum rules for
   divisibility by~11 in natural language, miscounts valid arrangements,
   and returns the wrong $N$. The code tool is never dispatched.}
\end{qablock}

\paragraph{Query corruption (thinking-induced misdirection).}
Thinking activates a strongly held prior about a sub-problem, which
redirects the search query toward confirming that prior rather than
retrieving the actual answer. The no-thinking orchestrator issues a
neutral, direct query and succeeds in a single call.

\begin{qablock}{Qwen3-8B}%
  {What word did two authors quote in distaste in Midkiff's 2014 article
   on dragon depictions? \textnormal{\tiny(GAIA)}}
\qapass{Tools}{fluffy}%
  {Tool calls: web\_search $\times$1. Neutral query \textit{Emily Midkiff journal Hreidmar's sons that guarded his house} retrieves the
   quoted word directly.}
\qafail{Tools + Thinking}{monstrous}%
  {Tool calls: web\_search $\times$1. Thinking fixates on decoding the
   mythological journal-name clue (``Hreidmar's son who guarded his
   house'' $\Rightarrow$ Reginn / Fáfnir); biased query surfaces
   heroic-dragon context; orchestrator concludes ``monstrous.''}
\end{qablock}

\paragraph{Premature closure.}
Thinking reasons about whether a resource is reachable \emph{before}
attempting any retrieval. Concluding incorrectly that a versioned online
document is inaccessible, the orchestrator returns a disclaimer
instead of the required one-word answer, without ever dispatching a
search.

\begin{qablock}{Qwen3-8B}%
  {Surname of the equine veterinarian in the CK-12 chemistry exercises
   (LibreTexts, compiled 08/21/2023)? \textnormal{\tiny(GAIA)}}
\qapass{Tools}{Louvrier}%
  {Tool calls: web\_search $\times$1. Searches without meta-reasoning;
   finds the surname directly.}
\qafail{Tools + Thinking}{Unable to determine without access to the document}%
  {No tool calls. Thinking reasons: ``I can't browse the internet or
   access specific versioned files directly.'' Having concluded the
   document is unreachable, the orchestrator returns a disclaimer
   rather than attempting retrieval. Output contract (single surname)
   is violated.}
\end{qablock}

\end{document}